\documentclass{conm-p-l}

\usepackage{amsmath}
\usepackage{amsfonts}
\usepackage{graphicx}
\usepackage{amsthm}
\usepackage{cite}

\numberwithin{equation}{section} \numberwithin{figure}{section}

\newtheorem{theorem}{Theorem}[section]

\newtheorem{Definition}[theorem]{Definition}

\newtheorem{Remark}[theorem]{Remark}

\begin{document}

\title[Asymptotics of $L_p$-norms of Hermite polynomials and R\'enyi entropy]
{Asymptotics of $L_p$-norms of Hermite polynomials \\ and R\'enyi entropy of
Rydberg oscillator states}

\author{A. I. Aptekarev}
\address{Keldysh Institute for Applied Mathematics, Russian Academy of Sciences and Moscow State University, Moscow, Russia}
\email{aptekaa@keldysh.ru}

\author{J.S. Dehesa}
\address{Departamento de F\'{\i}sica At\'omica, Molecular y Nuclear, Universidad de Granada, Granada, Spain}
\address{Instituto ``Carlos I'' de F\'{\i}sica Te\'orica y Computacional, Universidad de Granada, Granada, Spain}
\email{dehesa@ugr.es}

\author{P. S\'anchez-Moreno}
\address{Departamento de Matem\'atica Aplicada, Universidad de Granada, Granada, Spain}
\address{Instituto ``Carlos I'' de F\'{\i}sica Te\'orica y Computacional, Universidad de Granada, Granada, Spain}
\email{pablos@ugr.es}

\author{D. N. Tulyakov}
\address{Keldysh Institute for Applied Mathematics, Russian Academy of Sciences and Moscow State University, Moscow, Russia}
\email{dnt@mail.nnov.ru}



\begin{abstract}
The asymptotics of the weighted $L_{p}$-norms of Hermite polynomials, which
describes the R\'enyi entropy of order $p$ of the associated quantum oscillator
probability density, is determined for $n\to\infty$ and $p>0$. Then, it is
applied to the calculation of the R\'enyi entropy of the quantum-mechanical
probability density of the highly-excited (Rydberg) states of the isotropic
oscillator.
\end{abstract}

\maketitle

\section{Introduction}

A long standing problem in classical analysis and approximation theory is
the determination of the weighted $L_p$-norms
\begin{equation}
  \|\rho_n\|_p
    \equiv
    \left\{
    \int_\Delta \left[ \rho_n(x) \right]^p dx
    \right\}^\frac{1}{p}
  = \left\{
  \int_\Delta \left[ \omega(x) y_n^2(x) \right]^p dx
  \right\}^\frac{1}{p}
  ;\; p> 0,
  \label{eq:lp_norm}
\end{equation}
where $\{y_n(x)\}$ denotes a sequence of real polynomials orthogonal with
respect to the weight function $\omega(x)$ on the interval $\Delta$ so that
\[
  \int_{\Delta} y_n(x) y_m(x) \omega(x) dx =\delta_{m,n};\; m,n\in\mathbb{N},
\]
and
\begin{equation}
  \rho_n(x)=\omega(x) y_n^2(x).
  \label{eq:rakhmanov_density}
\end{equation}
We call (\ref{eq:rakhmanov_density}) Rakhmanov's probability density of the polynomial $y_n(x)$ since
this mathematician discovered in 1997 (see \cite{rakhmanov_mus77}) that it governs the asymptotic
$(n\to\infty)$ behaviour of the ratio $y_{n+1}/y_n$ for general $\omega > 0 \,\,\, a.
e.$ (positive almost everywhere) on the finite interval $\Delta$.
Physically,
$\rho_n(x)$ describes the radial probability density of the ground and excited
states of the physical systems whose non-relativistic wavefunctions are
controlled by the polynomials $y_n(x)$ \cite{galindo_pascual_90}.

The $L_p$-norms (\ref{eq:lp_norm})
are closely related to the frequency or entropic moments
\cite{zygmund_02,romera_02,grendar_04}
\[
W_p[\rho_n]=\left\langle \left[ \rho_n(x) \right]^{p-1}\right\rangle
=\int_\Delta \left[ \rho_n(x) \right]^p dx
=\| \rho_n\|_p^p,
\]
the R\'enyi entropies \cite{renyi_70}
\begin{equation}
R_p[\rho_n] = \frac{1}{1-p} \ln W_{p-1}[\rho];\; p>0,p\neq 1,
\label{eq:renyi_definition}
\end{equation}
and the R\'enyi spreading lengths \cite{hall_pra99}
\[
L_p^R[\rho_n]=\exp\left( R_p[\rho_n]
\right)=\|\rho_n\|_p^{-\frac{p}{p-1}}.
\]

These quantities have been applied in numerous fields from economics and
electrical engineering to chemistry, quantum physics and approximation
theory, as summarized in e.g. Refs
\cite{jizba_ap04,bialynicki_rudnicki,dehesa_caot11,dehesa_10,brody_pla07}.
The study of the $L_p$ norms of orthogonal polynomials is of independent
interest in the theory of general orthogonal and extremal polynomials.
This problem is connected with the
classical research of S.N.~Bernstein on the asymptotics of the $L_p$
extremal polynomials in \cite{Ber}, that has recently received further development \cite{Kal}, \cite{LS}. On the other hand, its statement is a
generalization of a widely known problem of Steklov on the estimation of the $\
L^\infty\ $ norms of polynomials orthonormal with respect to a positive weight
(see \cite{Su}). Indeed, for $p=1$ the norms are bounded (they are just equal to $1$);
however, for $p=\infty$ (as it has been shown by Rakhmanov \cite{Ra}) they may
grow to infinity. What happens with the boundedness of the $L_p$-norms of the
Rakhmanov density of the orthogonal polynomials when $1 \,<\, p\, < \, \infty $ ?
Our work sheds light on this issue for Hermite polynomials.

Recently these quantities have been calculated for polynomials $y_n(x)$
with arbitrary degree $n$ by means of the combinatorics-based Bell polynomials in the Hermite \cite{sanchezmoreno_jcam10}, Laguerre
\cite{sanchezmoreno_jcam11} and Jacobi \cite{guerrero_jpa10}
cases; see also \cite{dehesa_caot11}. However, this methodology is computationally very
demanding and analytically inefficient for high and very high values of
$n$.

The aim of this work is the asymptotic ($n\to\infty$) determination of the entropic moments
of Hermite polynomials, i.e.
\begin{equation}
W_p[\rho_n]=\int_{-\infty}^{+\infty}
\left[ e^{-x^2} H_n^2(x) \right]^p dx;\quad p> 0.
\label{eq:entropic momentH}
\end{equation}
The solution of this problem is relevant not only  \textit{per se} because it extends previous results \cite{aptekarev_rassm95,aptekarev_dm96,aptekarev_jcam10} obtained when $p\in\left[0,\frac43\right]$, but also because it paves the way for the evaluation of the $p$th-order R\'enyi entropy of the highly-excited (i.e., Rydberg) states of the physical systems whose radial wavefunctions are controlled by Hermite's polynomials such as, e.g. the oscillator-like systems.

The structure of the paper is the following. In Section \ref{sec2}, the
asymptotics $L_p$-norms of the Hermite polynomials is found for $p>0$. In
Section \ref{sec3}, these results are applied to evaluate the R\'enyi
entropy of Rydberg states of the quantum harmonic oscillator.
Finally, some
conclusions are given.

\section{$L_p$-norms of Hermite polynomials: Asymptotics ($n\to\infty$)}
\label{sec2}

In this section we find the main term of asymptotics $W_p[\rho_{n-1}]$
(see (\ref{eq:entropic momentH})) when $n \rightarrow \infty$. Here we consider
Hermite polynomials $H_n(x)$ in the standard normalization
\begin{equation}
  H_n(x)=(2x)^n \,_2F_0\left(
  \begin{array}{c}
  -\frac{n}{2}, \frac{1-n}{2}\\
  -
  \end{array}
  ;-\frac{1}{x^2}
  \right).
  \label{eq:hermite_normalization}
\end{equation}
They satisfy the recurrence relation
\[
H_{n+1}(x)=2 x H_n(x)-2n H_{n-1}(x),\quad
H_0(x)=1, \quad H_{-1}(x)=0,
\]
and have the following norm
\begin{equation}
  h_n=\int_{-\infty}^{+\infty}\left[  H_n^2(x) \, e^{-x^2}\right]dx\,=\, \sqrt{\pi}\, n!\,\, 2^n \,.
  \label{eq:hermite_norm}
\end{equation}

To reach the goal we need to have good asymptotics for $H_n(x)$ on
$\mathbb{R}$. The first strong asymptotics formulae for Hermite
polynomials are due to Plancherel and Rotach \cite{plancherel_cmh29}
(see also \cite{szego_75}). They
describe polynomials $H_n$ when $n\to\infty$ in the following subdomain of
$\mathbb{R}$:
\begin{equation}
\begin{array}{lll}
  {\rm a) } & x=\sqrt{2n+1} \cosh y, \; & \epsilon \le y <\infty;\\
  {\rm b) } & x=\sqrt{2n+1} +n^{-\frac16} t,\; &  t\in K\subset \mathbb{C};\\
  {\rm c) } & x=\sqrt{2n+1}\cos y,\; &\epsilon \le y\le \pi-\epsilon.
\end{array}
  \label{eq:plancherel_cases}
\end{equation}
Using Plancherel--Rotach formulae (and their generalization for Freud weights
from \cite{rakhmanov_92}) it was obtained \cite{aptekarev_rassm95} for the orthonormal
Hermite polynomials $\tilde{H}_{n-1}(x)$ that
\begin{equation}
  \int_{-\infty}^{+\infty} \left[ \tilde{H}_{n-1}^2(x) e^{-x^2} \right]^p dx
  =\,
  c_p
  (2n)^{\frac{1-p}{2}} \left(1+o(1)\right)\,,\quad \mbox{for}\,\, p\le
\frac43\,,
  \label{eq:plancherel_result}
\end{equation}
where
\begin{equation}
c_p\,=\, \left(\displaystyle\frac{2}{\pi}\right)^{p}\,\displaystyle\frac{\Gamma(p+\frac{1}{2})}{\Gamma(p+1)}\,
\,\displaystyle\frac{\Gamma(1-\frac{p}{2})}{\Gamma(\frac{3}{2}-\frac{p}{2})}\,.
  \label{eq:cp}
\end{equation}
Using (\ref{eq:hermite_norm}) and the Stirling's formula, this expression together with (\ref{eq:entropic momentH}) produces the following asymptotics for the entropic moment of the orthogonal Hermite polynomial $H_{n-1}(x)$:
\begin{equation}
 W_p[\rho_{n-1}]
=c_p\,\,h_{n-1}^p\,\,(2n)^{\frac{1-p}{2}} \,\left(1+o(1)\right)
=c_p
\pi^p(2n)^{p(n-1)+1/2}\, e^{-pn}\,\left(1+o(1)\right)\,.
  \label{eq:plancherel_resultST}
\end{equation}
The restriction on $p$ in (\ref{eq:plancherel_result}) appeared because
the Plancherel--Rotach formulae in (\ref{eq:plancherel_cases}) do not match each other, i.e. subdomains in
(\ref{eq:plancherel_cases}) do not intersect. Particularly, there is a gap
between zone a) and zone b) in (\ref{eq:plancherel_cases}), which plays an
important role for the limit (\ref{eq:entropic momentH}) when $n\rightarrow\infty$.
In the asymptotics of
(\ref{eq:plancherel_result}) the main contribution in the left hand side
integral gives the part of the integral described in a) of
(\ref{eq:plancherel_cases}). The gap between zone a) and zone b) gives the
main contribution in the integral for bigger $p$.

The asymptotic description of Hermite polynomials in the subdomains
covering all $\mathbb{R}$ was obtained not so long ago. In 1999
Deift et al \cite{deift_cpam99} (see also \cite{deift_99})
have
obtained the global asymptotic portrait of polynomials orthogonal with
respect to exponential weights by means of the powerfull matrix Riemann-Hilbert method. As a collorary for Hermite polynomials
$H_n(\sqrt{2n}z)$, they obtained asymptotics as $n\to\infty$ and $z$
belongs to
\begin{equation}
\begin{array}{ll}
  {\rm a) } & |z|\ge 1+\delta;\\
  {\rm b) } & 1-\delta \le |z| \le 1+\delta;\\
  {\rm c) } & |z|\le 1-\delta,
\end{array}
  \label{eq:deift_cases}
\end{equation}
for small $\delta>0$.
Evidently, there are no gaps between zones and Deift et al's asymptotics can be
used for obtaining asymptotics of (\ref{eq:entropic momentH}) for bigger $p$.

Recently a new approach for obtaining the global asymptotic
portrait of orthogonal polynomials has appeared \cite{tulyakov_rassm10}. Contrary to the matrix
Riemann-Hilbert method which starts from the weight of orthogonality, the
starting point in \cite{tulyakov_rassm10} is the recurrence relation
which characterizes the orthogonal polynomials. The application of this approach to
Hermite polynomials brought an asymptotic description in the following
subdomains of $\mathbb{R}$:
\begin{equation}
  \begin{array}{ll}
  {\rm a) } & x^2\in[2n+n^{\frac13+\theta};\infty);\\
  {\rm b) } & x^2\in[2n-n^{\frac13+\theta};2n+n^{\frac13+\theta}];\\
  {\rm c) } & x^2\in[0;2n-n^{\frac13+\theta}].
\end{array}
  \label{eq:tulyakov_cases}
\end{equation}
for $\theta\in(0;\frac23)$. It is worth noting that zones a) and c) in (\ref{eq:tulyakov_cases}) are wider than
zones a) and c) in (\ref{eq:deift_cases}); in its turn zone b) in
(\ref{eq:deift_cases}) is wider that b) in (\ref{eq:tulyakov_cases}). In
these zones we take $\theta<\frac16$ for Hermite polynomials.
Then, it follows from Theorem 5 of \cite{tulyakov_rassm10} that:
\begin{equation}
  \begin{array}[]{ll}
   {\rm in\, a) }\;  \displaystyle
    H_{n-1}(x)=\frac{1}{\sqrt{2}}
    \frac{\left( x+\sqrt{x^2-2n}\right)^{n-\frac12}}{\sqrt[4]{x^2-2n}}
    \exp\left( \frac{x^2-n-x \sqrt{x^2-2n}}{2} \right)
    \left( 1+o(1) \right),\\
     {\rm in\, c) } \; \displaystyle
    H_{n-1}(x)= \sqrt{2}
    \frac{\left( \sqrt{2n} \right)^{n-\frac12}}{\sqrt[4]{2n-x^2}}
    \exp\left( \frac{x^2-n}{2} \right)\\
      \displaystyle \times
    \cos\left( \left( n-\frac12 \right) \arcsin\left(
    \sqrt{1-\frac{x^2}{2n}} \right)
    -\frac{x \sqrt{2n-x^2}}{2}-\frac{\pi}{4}
    +o(1)\right)
    \left( 1+o(1) \right),\\
     {\rm in\, b) } \; \displaystyle
    H_{n-1}(x)=\frac{\sqrt{2\pi}}{\sqrt[6]{2}}
    x^{n-\frac23}
    \exp\left( \frac{x^2}{4}+o(1) \right)
    {\rm Ai}\left( -\frac{\sqrt[3]{2}}{2}z+o\left( n^{-\frac13} \right) \right)
    \left( 1+o(1) \right),
  \end{array}
  \label{eq:hermite_asymptotics}
\end{equation}
where $ z:=\frac{2n}{x^\frac23}-x^\frac43,$ and ${\rm Ai}$ denotes the Airy function (see \cite{abramowitz_72}, page 367).

Using these asymptotics we can obtain the main result of this section.
\begin{theorem}
Let $H_n(x)$ be the Hermite polynomials with the standard normalization (\ref{eq:hermite_normalization}).
Then the frequency or entropic moments $W_p[\rho_{n-1}]$, given by Eq. (\ref{eq:entropic momentH}), have for
$n\to\infty$ the following asymptotic values
\begin{equation}
  W_p[\rho_{n-1}]
  =\left\{
  \begin{array}[]{ll}
    c_p
    \pi^p \,(2n)^{p(n-1)+1/2}\, e^{-pn}\,\left(1+o(1)\right),\quad &\, p<2,\\[0.3em]
    2(2n)^{2n-\frac32} e^{-2n} \left( \ln(n)+O(1) \right),\quad &\,    p=2,\\[0.3em]
   2C_p \,2^{-p}\,(2n)^{p\left( n-\frac23 \right)-\frac16} e^{-pn}
    \left( 1+o(1) \right),\quad &\, p>2.
  \end{array}
  \right.
  \label{eq:goal_limit}
\end{equation}
where the constant $c_p$ is defined in (\ref{eq:cp}) and the constant $C_p$ is equal to
\[
C_p=\int_{-\infty}^{+\infty}
\left[
\frac{2\pi}{\sqrt[3]{2}}\,\, {\rm Ai}^2\left(
-\frac{z \sqrt[3]{2}}{2}
\right) \right]^p dz\,.
\]
\label{theorem}
\end{theorem}

We note that the first asymptotic formula in the right hand side of
(\ref{eq:goal_limit}) coincides with (\ref{eq:plancherel_resultST}), but now it
holds true in the maximal range of $p$ (when $p=2$, then $c_p=\infty$);
let us also highlight that  the main
term of the asymptotics is growing. Moreover, the smaller terms contain a constant
which depends on $p$, and when  $p\to 0$ this constant tends to infinity; however,
our formula is correct for any small fixed $p>0$.
We also note
that the leading term of all three formulae in the right hand side of
(\ref{eq:goal_limit}) match each other when $p\rightarrow 2$.

\begin{proof}
 Doing identical
transformations and some evident asymptotic estimates, we have from (\ref{eq:hermite_asymptotics}) that:
\begin{equation}
  \begin{array}[]{ll}
     {\rm in\, a) } \; \displaystyle
    H_{n-1}^2(x) e^{-x^2} =
    \frac{(2n)^{n-1}}{2}e^{-n}
    \exp\left[
    (2n-1) {\rm arccosh} \frac{x}{\sqrt{2n}}
    -x \sqrt{x^2-2n} \right]\\
    \times
    \left( \frac{x^2}{2n}-1 \right)^{-\frac12}
    \left( 1+o(1) \right);\\
     {\rm in\, c) } \; \displaystyle
    H_{n-1}^2(x) e^{-x^2} =
    (2n)^{n-1} e^{-n}\\
    \times
    \left\{
    1-\sin\left[ (2n-1)\arcsin \sqrt{1-\frac{x^2}{2n}}
    -x \sqrt{2n-x^2}+o(1)
    \right] \right\}\\
    \times \left( 1-\frac{x^2}{2n} \right)^{-\frac12}
    \left( 1+o(1) \right);\\
    {\rm in\, b) } \; \displaystyle
    H_{n-1}^2(x) e^{-x^2} =
    (2n)^{n-\frac23} e^{-n} \frac{2\pi}{\sqrt[3]{2}}
    \left( \frac{x}{\sqrt{2n}} \right)^{2\left(n-\frac23\right)}
    \exp \left( \frac{2n-x^2}{2} \right)\\
    \times
    {\rm Ai}^2\left( -\frac{\sqrt[3]{2}}{2}z+o\left( n^{-\frac13} \right)
    \right)
    \left( 1+o(1) \right).
  \end{array}
  \label{eq:density_asymptotics}
\end{equation}

Now we start to estimate the integral in (\ref{eq:entropic momentH}). We consider the
interval of integration $[0,\infty)$ (since the integral is even) and split
it in the subintervals a), b), c) as in (\ref{eq:tulyakov_cases}). And we split the interval b) in (\ref{eq:tulyakov_cases}) into three
subintervals:
\begin{equation}
  \begin{array}[]{ll}
    {\rm b}_1) & x^2 \in
    \left[ 2n-n^{\frac13+\theta}; 2n-Mn^\frac13 \right];\\[3mm]
    {\rm b}_2) & x^2 \in
    \left[ 2n-Mn^\frac13; 2n+Mn^\frac13 \right];\\[3mm]
    {\rm b}_3) & x^2 \in
    \left[ 2n+Mn^\frac13; 2n+n^{\frac13+\theta} \right].
  \end{array}
  \label{eq:tulyakov_subb}
\end{equation}
Thus, we have splitted $x\in[0,\infty)$ on five zones (see Figure
\ref{fig:tulyakov_subb}).

\begin{figure}[ht]
  \begin{center}
   \includegraphics[width=12.5cm]{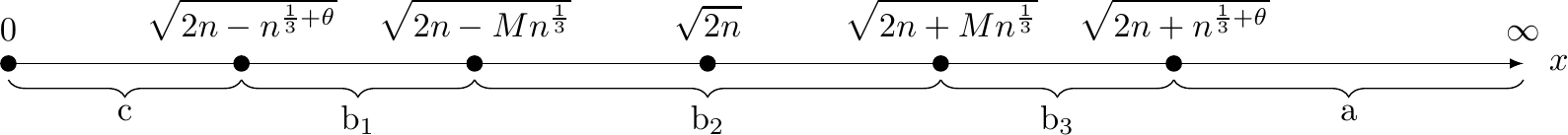}
  \end{center}
  \caption{Zones of $\mathbb{R}_+$, which gives different contribution to
  the integral, depending on $p$.}
  \label{fig:tulyakov_subb}
\end{figure}

We recall, that $\theta$ is a fixed small number, such that
$0<\theta<\frac16$ and the constant $M$ will be chosen depending on $p$.

Making the change of variables $\frac{x}{\sqrt{2n}}=t$ in
(\ref{eq:density_asymptotics}), we obtain for the integrals along the interval
a) in (\ref{eq:tulyakov_cases}):
\begin{eqnarray}
   I_a&=&\int\limits_{\sqrt{2n+n^{\frac13+\theta}}}^\infty
  \left( H_{n-1}^2(x) e^{-x^2} \right)^pdx\nonumber 
   =
  (2n)^{p(n-1)+\frac12} e^{-pn} 2^{-p}\\
   &&\times 
  \int\limits_{1+\frac14n^{\theta-\frac23}+\epsilon_n}^\infty
  \exp\left[ p(2n-1){\rm arccosh}\,t-2ntp\sqrt{t^2-1}+o(1) \right]
  \frac{dt}{\left(t^2-1\right)^\frac{p}{2}},
  \label{eq:integral_a}
\end{eqnarray}
and for the interval c) in (\ref{eq:tulyakov_cases}):
\begin{align*}
   &I_c=\int_0^{\sqrt{2n-n^{\frac13+\theta}}}
  \left( H_{n-1}^2(x) e^{-x^2} \right)^p dx
   =
  (2n)^{p(n-1)+\frac12} e^{-pn}
  \nonumber \\
  &\times
  \int\limits_0^{1-\frac14n^{\theta-\frac23+\epsilon_n}}
  \left[ 1-\sin\left( (2n-1)\arcsin\sqrt{1-t^2}
  -2nt\sqrt{1-t^2}\right) +o(1) \right]^p \frac{dt}{(1-t^2)^\frac{p}{2}}\nonumber \\
  &=
  (2n)^{p(n-1)+\frac12} e^{-pn}
  \nonumber \\
   &\times
  \int\limits_0^{1-\frac14n^{\theta-\frac23+\epsilon_n}} 2^p
  \sin^{2p}
  \left(
  \frac{(2n-1)\arcsin\sqrt{1-t^2}-2nt\sqrt{1-t^2}}{2}-\frac{\pi}{4} +o(1)
  \right)
  \frac{dt}{(1-t^2)^\frac{p}{2}},
\end{align*}
where $\epsilon_n=o\left( n^{\theta-\frac23} \right)$.

Then we pass to the integrals along b)-(\ref{eq:tulyakov_cases}). The idea
to split interval b) into three subintervals (\ref{eq:tulyakov_subb}) was
because we are intending to use in the subintervals b$_1$) and b$_3$) in
(\ref{eq:tulyakov_subb}) the asymptotics of the Airy function from
b)-(\ref{eq:density_asymptotics}); in
b$_2$)-(\ref{eq:tulyakov_subb}) we shall use the explicit expression for the
Airy function. Noticing that for $n\to\infty$, we have from definition of
$z$ in (\ref{eq:hermite_asymptotics})
\[
z=\frac{2n}{x^{\frac23}}-x^\frac43
\Rightarrow
x\simeq \sqrt{2n}-\frac{z}{2 \sqrt[6]{2n}}
\Rightarrow
dx=-\frac{dz}{2 \sqrt[6]{2n}}
\]
Thus,
\begin{eqnarray*}
I_{{\rm b}_1} &=& \int\limits_{\sqrt{2n-n^{\frac13+\theta}}}^{\sqrt{2n-Mn^\frac13}}
\left( H_{n-1}^2(x) e^{-x^2} \right)^p dx\nonumber \\
&\simeq&
(2n)^{p\left( n-\frac23 \right)-\frac16} e^{-pn} \frac12
\int\limits_M^{n^\theta} \left[ 1+\sin\left( \frac23 z^\frac32 \right) \right]^p
z^{-\frac{p}{2}} dz,
\end{eqnarray*}
\begin{eqnarray}
I_{{\rm b}_3}&=&\int\limits_{\sqrt{2n+Mn^\frac13}}^{\sqrt{2n+n^{\frac13+\theta}}}
\left( H_{n-1}^2(x) e^{-x^2} \right)^p dx\nonumber\\
&\simeq&
(2n)^{p\left( n-\frac23 \right)-\frac16} e^{-pn} 2^{-p-1}
\int\limits_M^{n^\theta} \exp\left( -\frac23 p z^\frac32 \right)z^{-\frac{p}{2}}
dx;
\label{eq:integral_b3}
\end{eqnarray}
\begin{eqnarray}
I_{{\rm b}_2}&=&\int\limits_{\sqrt{2n-Mn^\frac13}}^{\sqrt{2n+Mn^\frac13}}
\left( H_{n-1}^2(x) e^{-x^2} \right)^pdx\nonumber\\
&\simeq&
(2n)^{p\left( n-\frac23 \right)-\frac16} e^{-pn} 2^{-p-1}
\int\limits_{-M}^M \left[ \frac{2\pi}{\sqrt[3]{2}} {\rm Ai}^2\left(
-\frac{z\sqrt[3]{2}}{2} \right) \right]^p dz.
\label{eq:integral_b2}
\end{eqnarray}
The symbol $\simeq$ means that the ratio of the left and right hand sides tends to unity.
Now we can analyse the contributions of the various $p$-depending parts of the integral of the left hand side of (\ref{eq:goal_limit}), when $n\to\infty$.
We note, that all $o(1)$ terms in our asymptotic analysis are differentiable,
therefore they will not make contributions in our further estimates of the
integrals.

First, we notice that the integral part of $I_{\rm a}$ in the right hand
side of (\ref{eq:integral_a}) is exponentially small, and there exist
constants $\alpha,c>0$, such that this integral is estimated as
$O\left(n^\alpha
\exp\left[ -c n^{\frac32\theta} \right]\right)$, and for $I_{\rm a}$ we
have
\[
I_{\rm a}=\frac{(2n)^{p(n-1)+\frac12}}{2^p e^{np}}
O\left( n^\alpha \exp\left[ -c n^{\frac32 \theta} \right] \right).
\]
Therefore this part is negligible for (\ref{eq:goal_limit}).

Second, we notice that the integral parts of $I_{ {\rm b}_3}$ and $I_{ {\rm
b}_2}$ in the right hand side of (\ref{eq:integral_b3}) and
(\ref{eq:integral_b2}), respectively, are $O(1)$ and we have
\[
I_{ {\rm b}_2}, I_{ {\rm b}_3} =(2n)^{p\left(n-\frac23\right)-\frac16}
e^{-pn} 2^{-p-1} O(1).
\]

When $p<2$, the integral parts of $I_{ {\rm c}}$ and $I_{ {\rm b}_1}$
behave as $O\left( 1 \right)$ and $O\left( n^{\theta
\frac{2-p}{2}} \right)$ and we have
\[
I_{\rm c}=(2n)^{p(n-1)+\frac12} e^{-pn} O\left( 1 \right),
\]
\[
I_{{\rm b}_1}=(2n)^{p\left( n-\frac23 \right)-\frac16}
\frac{e^{-pn}}{2} O\left( n^{\theta \frac{2-p}{2}} \right),
\]
for $0<p<2$. Therefore, when $p<2$, only $I_{\rm c}$ gives contribution in
(\ref{eq:goal_limit}).
Thus, we have proved, that (\ref{eq:plancherel_result}) is valid for
$0<p<2$.

When $p=2$, then both integral parts $I_{\rm c}$ and $I_{ {\rm b}_1}$ have
the same logarithmic rate of growth $O(\ln n)$. Computing the constant in
$O$ we obtain in a non-trivial way that
\[
\int_{0}^\infty \left( H_{n-1}^2(x) e^{-x^2} \right)^2dx
=(2n)^{2n-\frac32} e^{-2n} \left( \ln(n)+O(1) \right),\; {\rm for}\; p=2.
\]

Finally, for $p>2$ as we see from
(\ref{eq:integral_a})-(\ref{eq:integral_b2}), the integral over
b)-(\ref{eq:tulyakov_cases}) dominates in (\ref{eq:goal_limit}).
Thus taking $M\to\infty$, we obtain
\begin{eqnarray*}
 \int_{-\infty}^\infty
\left[ H_{n-1}^2(x) e^{-x^2} \right]^p dx\\
 =
(2n)^{p\left( n-\frac23 \right)-\frac16} e^{-pn} 2^{-p}
\left(
\int_{-\infty}^\infty \left[
\frac{2\pi}{\sqrt[3]{2}} {\rm Ai}^2\left(
-\frac{z \sqrt[3]{2}}{2}
\right) \right]^p dz \right)
\left( 1+o(1) \right);\; p>2.
\end{eqnarray*}

\end{proof}

\section{R\'enyi entropy of Rydberg oscillator states}
\label{sec3}

In this section Theorem \ref{theorem} is applied to obtain the R\'enyi entropy of the Rydberg states of the one-dimensional harmonic oscillator, described by the quantum-mechanical potential $V(x)=\frac12 x^2$.
It is in this energetic region where the transition from classical to quantum correspondence takes place.
The physical solutions of the Schr\"odinger equation for the harmonic oscillator system (see e.g., \cite{galindo_pascual_90}), are given by the wavefunctions characterized by the energies  $E_n=n+\frac12$ and the quantum probability densities
\[
\tilde{\rho}_n(x) = \frac{1}{\sqrt{\pi}n! 2^n} e^{-x^2} H_n^2(x) \equiv e^{-x^2} \tilde{H}_n^2(x),
\]
where $\tilde{H}_n(x)$ denotes the orthonormal Hermite polynomials of degree $n$. The degree $n=0,1,2,\ldots$ labels the energetic level.

The entropic moments of these densities $W_p[\rho_n]$ are expressed in terms of the entropic moments of the Hermite polynomials  as
\begin{equation}
W_p[\tilde{\rho}_n]=\frac{1}{\pi^{\frac{p}{2}}(n!)^p 2^{pn}}W_p[\rho_n].
\label{eq:rho_to_H}
\end{equation}
Thus, the entropic moments of the harmonic oscillator states are given by the entropic moments of the orthonormal Hermite polynomials. Consequently, according to equation (\ref{eq:renyi_definition}) the R\'enyi entropy of the harmonic oscillator for both ground and excited states is given by the concomitant R\'enyi entropy of the involved orthonormal Hermite polynomials.

Let us now consider the Rydberg states of the oscillator system; that is, the states with high and very high values of $n$. Then, taking into account equations (\ref{eq:goal_limit}) and (\ref{eq:rho_to_H}), we obtain the asymptotic ($n\to\infty$) values 
\begin{equation}
  W_p[\tilde{\rho}_{n-1}]
 =\left\{
  \begin{array}[]{ll}
    c_p \, (2n)^{\frac{1-p}{2}}\,\left(1+o(1)\right),
    \quad &\, p<2,\\[0.3em]
    \displaystyle
    2\pi^{-2} (2n)^{-\frac12} \left(\ln(n)+O(1)\right) 
    ,\quad &\,
    p=2,\\[0.3em]
   \displaystyle 2\frac{C_p}{(2  \pi)^p} \,(2n)^{-\frac16(p+1)}
    \left( 1+o(1) \right),\quad &\, p>2,
  \end{array}
  \right.
  \label{eq:goal_limitNORM}
\end{equation}
for the entropic moments of the Rydberg oscillator states. Finally, it is straightforward to have the expressions for the R\'enyi entropy of the Rydberg states as follows from Eqs. (\ref{eq:goal_limitNORM}) and (\ref{eq:renyi_definition}).

\begin{figure}
\begin{center}
\includegraphics[width=10cm]{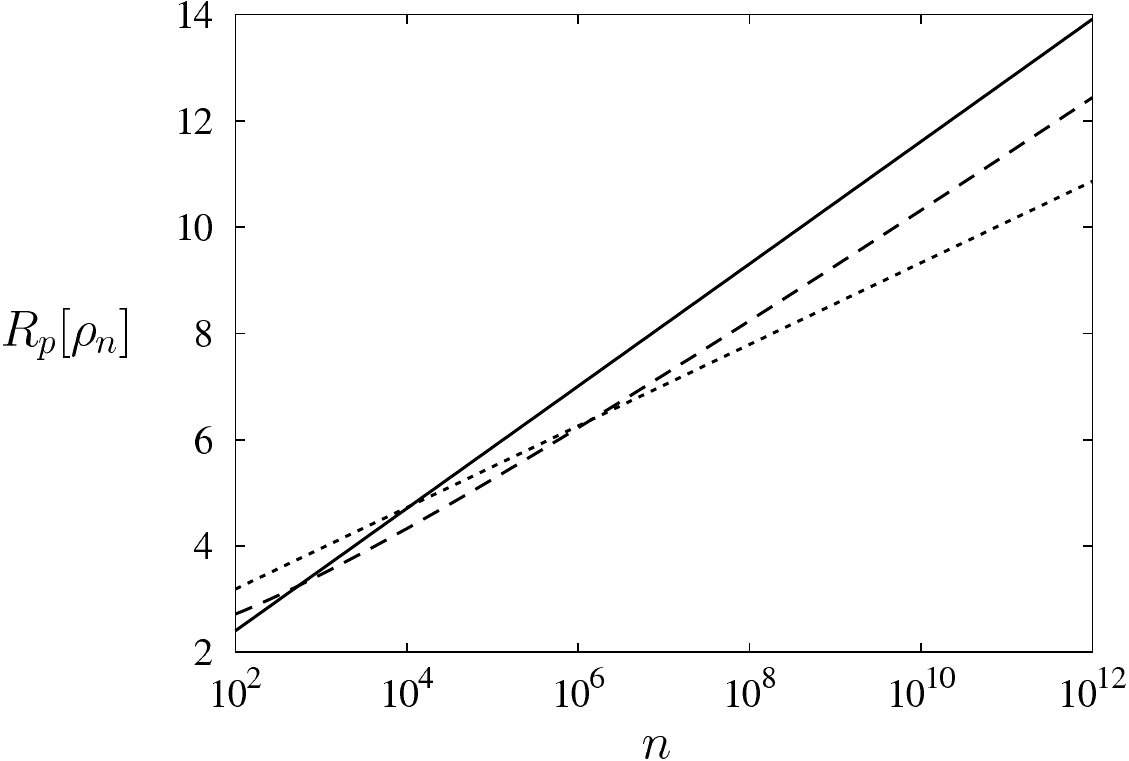}
\end{center}
\caption{R\'enyi entropy $R_p[\rho_n]$ for $p=\frac32$ (solid line), $p=2$ (dashed line) and $p=3$ (dotted line) of the Rydberg oscillator states with $n=100$ to $n=10^{12}$.}
\label{fig2}
\end{figure}

Figure \ref{fig2} shows the values of the R\'enyi entropy $R_p[\rho_n]$ for $p=\frac32$, $p=2$ and $p=3$, as a function of $n$ from $n=100$ to $n=10^{12}$. Notice that in all the cases the R\'enyi entropy increases with $n$. This indicates that the spreading of these states increases with $n$. Moreover, for the values of $p$ considered, after some initial intersections, the R\'enyi entropy also increases when $p$ decreases, for very large values of $n$, ($n > 10^7$). This is also the observed behaviour when the R\'enyi entropy is exactly calculated for low and moderate values of $n$ (see e.g. \cite{sanchezmoreno_jcam10}). Then, we can conclude that the observed intersections come from the differences between the asymptotic and the exact values of this quantity.

\section{Conclusions}
\label{conclusions}

In this work, we have shown that the R\'enyi entropy of the one-dimensional harmonic oscillator is exactly equal to the R\'enyi entropic integral of the involved orthonormal Hermite polynomials. Then,
we have calculated the R\'enyi entropy of the highly excited states of the oscillator system by use of the asymptotics ($n\to \infty$) of the $L_p$-norms of the Hermite polynomials $H_n(x)$ which control the corresponding wavefunctions. Remark that no recourse to the quasi-classical approximation has been done.  The asymptotics of the $L_p$-norms of $H_n(x)$ was determined by extending some sophisticated ideas and techniques extracted from the modern approximation theory \cite{aptekarev_rassm95,tulyakov_rassm10}. This research opens the way to investigate the asymptotics of the multivariate Hermite polynomials, what would allows one to compute the R\'enyi entropy of the Rydberg states of the harmonic oscillator of arbitrary dimensionality.

\section*{Acknowledgements}

AIA and DT are partially supported by the grant RFBR 11-01-12045 OFIM.
AIA is partially supported by the grant RFBR 11-01-00245 and the Chair Excellence Program of Universidad Carlos III Madrid, Spain and Bank Santander.
DT are partially supported by the grant RFBR 10-01-00682.
JSD and PSM are very grateful for partial support to Junta de Andaluc\'{\i}a
(under grants FQM-4643 and FQM-2445) and Ministerio de Ciencia e Innovaci\'on under project FIS2011-24540. JSD and PSM belong to the Andalusian research group FQM-0207.

\bibliographystyle{amsplain}
\bibliography{renyi_hermite}

\end{document}